\documentclass[
reprint,
floatfix,
longbibliography,
]{revtex4-1}

\usepackage{xcolor}
\usepackage[colorlinks=true,citecolor=blue,linkcolor=magenta]{hyperref}
\usepackage{graphicx}
\usepackage{bm,amsmath,amssymb}

\begin{document}

\author{Bin Yan}
\affiliation{Theoretical Division, Los Alamos National Laboratory, Los Alamos, New Mexico 87545, USA}

\title{Bell meets Cavendish: a quantum signature of gravity?}

\begin{abstract}
The inclusion of gravitation within the framework of quantum theory remains one of the most prominent open problem in physics. To date, the absence of empirical evidence hampers conclusions regarding the fundamental nature of gravity---whether it adheres to quantum principles or remains a classical field manifests solely in the macroscopic domain. This article presents a though experiment aimed at discerning the quantum signature of gravity through the lens of macroscopic nonlocality. The experiment integrates a standard Bell test with a classical Cavendish experiment. We illustrate that the measurement apparatuses employed in a Bell experiment, despite lacking entanglement, defy classical descriptions; their statistical behaviors resist explanations through local hidden variable models. Extending this argument to encompass the massive objects in the Cavendish experiment allows for further disputing classical models of the gravitational field. Under favorable conditions and in light of corroborating evidence from the recent loophole-free Bell experiments, the quantum character of gravity is essentially substantiated.
\end{abstract}

\maketitle

Quantum theory and general relativity stands as foundational cornerstones of contemporary physics; nevertheless, their seamless integration has proven to be a persistent challenge. Identification of an unequivocal candidate for quantum gravity remains an elusive pursuit. More significantly, the question of whether gravity is inherently a quantum entity continues to be a subject of extensive debate~\cite{DeWitt1957,Penrose1996,Eppley1977,Marletto2017Why}. This unresolved issue has prompted investigations into alternative approaches that treat gravitational field as merely a classical background upon which quantum matter resides and interacts~\cite{Oppenheim2023,Oppenheim2023NatComm,Salcedo2012,Barcelo2012}.

The assessment of gravity's intrinsic quantum nature necessitates verification through experiments~\cite{Marketto2017}. Unfortunately, empirical evidences for the quantum aspects of gravity are still missing. On small scales where quantum effects predominate, gravity is exceedingly small: the gravitational coupling constant is smaller by about $43$ orders of magnitude than the fine structure constant. This renders the prospect of searching for gravitational effect on the microscopic level, such as detection of the hypothetical graviton~\cite{Rothman2006,Dyson2013}. In contrast, within the macroscopic domain, where gravity can dominate other forces, the observation of quantum signatures in massive objects remains absent. Possibilities exist for a breakdown of quantum mechanics at macroscopic scales where gravitational effects become prominent~\cite{Ghirardi1985,Diosi1989,Penrose1996,Pfister2016}. This poses formidable technical and conceptual challenges to proposals seeking to unveil quantum manifestations of gravity through macroscopic superposition or entanglement~\cite{Marshall2003,Isart2011,Mari2016,Marletto2017,Bose2017,Belenchia2018}. 

Is it possible to devise an experiment capable of probing the quantum signatures of gravity without necessitating quantum superposition of massive objects (if it is ever attainable)? Here, we argue that such an experiment exists and it may have already been, at least to some extent, realized.

\begin{figure}[b!]
    \centering
    \includegraphics[width=\columnwidth]{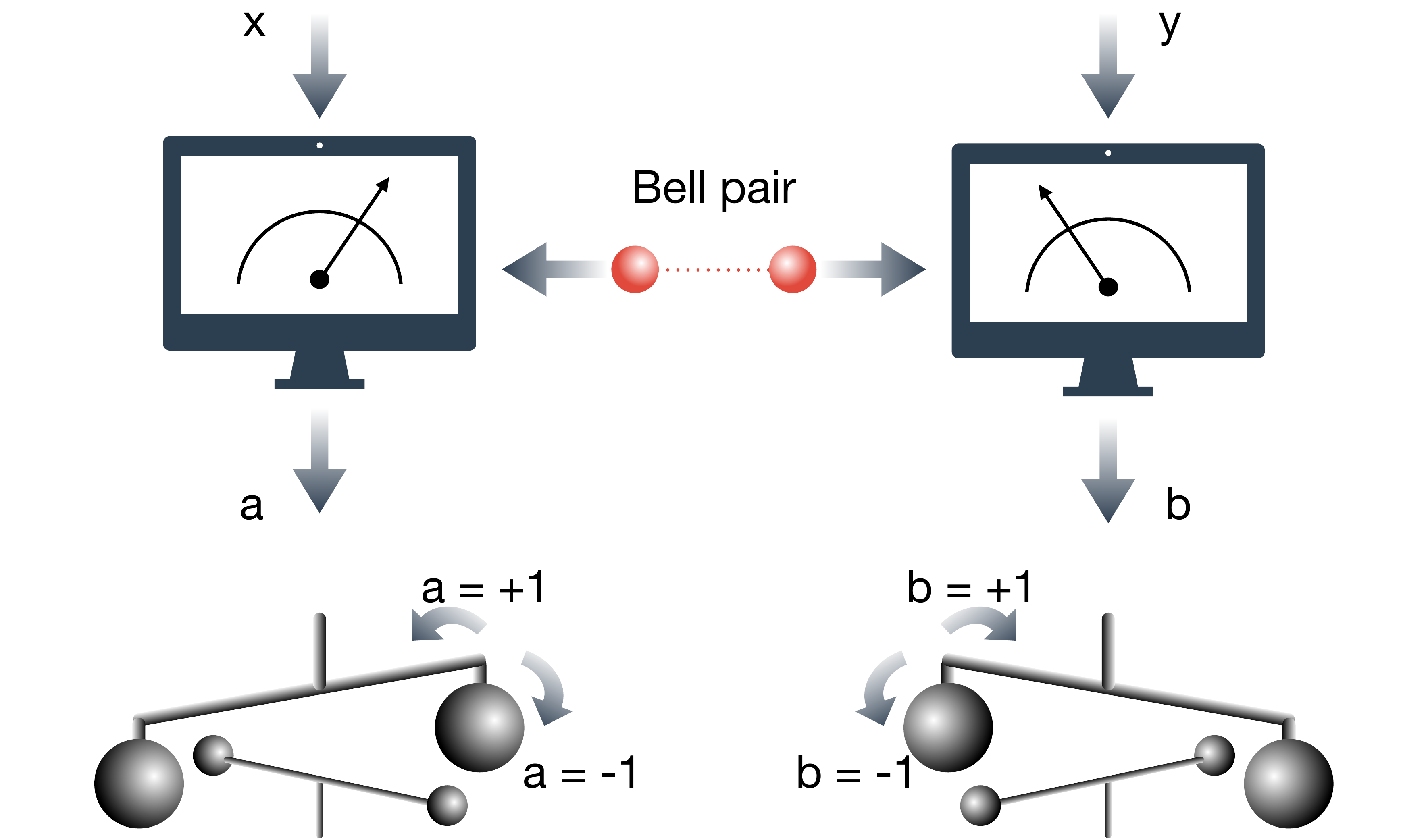}
    \caption{\textbf{Top: A standard Bell experiment}, where each party performs two binary measurements at will on a pair of entangled particles. The binary variables $x,y$ and $a,b$ denote respectively the measurement choices and outcomes. \textbf{Bottom: A classical Cavendish experiment}, where the outer larger spheres undergo rotation to distinct positions based on the measurement outcomes $a$ and $b$ from a preceding Bell experiment. The inner smaller balls, attracted to the larger spheres, function as detectors of the gravitational field produced by the larger spheres. Correlations between the positions of the inner balls cannot be reproduced by local deterministic models, thereby negating the classical interpretation of gravitational fields.}
    \label{fig:BellCavendish}
\end{figure}

\vspace{8pt}
Let us start by considering a standard Bell experiment~\cite{Freedman1971,Aspect1981,Aspect1982}, as inllustrated in figure~\ref{fig:BellCavendish}, wherein two parties are spatially separated by a spacelike distance. Each party select at will between two types of measurements, denoted by the binary variables $x$ and $y$. Each of these measurements yields two possible outcomes labeled by the binary variables $a$ and $b$. In classical probability theory, the local measurements are represented by response functions~\footnote{The response function $F(\xi)$ is a non-negative function bounded by unity. $F(\xi) = \{0,1\}$ are called indicator functions representing ideal deterministic measurements. Here we consider generic response functions which accommodate non-deterministic probability event. $F(\xi)$ can be interpreted as the probability for the corresponding event to happen, given the hidden variable $\xi$.} $F_x^a(\xi)$ and $G_y^b(\eta)$ on their respective local probability space. Here, the parameters $\xi$ and $\eta$ are conventionally interpreted as the unknown `hidden variables' that underlies the probabilistic behaviors of the systems. The probability for each outcome given a choice of measurement setting is produced by a joint distribution $\rho(\xi,\eta)$, i.e.,
\begin{equation}
    P(a,b|x,y) =  \int d\xi d\eta ~ \rho(\xi,\eta)F_x^a(\xi)G_y^b(\eta).
\end{equation}
Bell's theorem~\cite{Bell1964,Clauserf1969,Fine1982,Brunner2014} asserts that there exist local quantum measurements performed upon certain entangled states, whose statistics cannot be reproduced by classical probabilities of the above form. 

Let us redirect our focus to the measurement apparatuses in the Bell test, isolating themselves as the entities subject to measurement. In this context, the term `apparatus' denotes the macroscopic degree of freedom of the measurement devices, such as the spatial position of the physical pointers that signify the measurement outcomes, or a binary digital signal displayed on a monitor. The question at hand pertains to describing the apparatuses as classical objects, a notion may seem intuitive. However, their macroscopic status, e.g., pointer positions, are still probabilistic and statistical correlated. Such correlations between the two distant apparatuses may be resistant to classical interpretations as well (otherwise Bell's theorem will be invalidated).

Indeed, a formal and yet straightforward proof can be formulated as the following. Let us model the macroscopic outcomes by a different set of response functions $f_x^a(\xi)$ and $g_y^b(\eta)$. An additional layer of complexity arises due to the impact of the choices of measurement settings $x$ and $y$ on the apparatuses: The selection of a measurement setting necessitates an operation on the apparatues, involving actions such as tuning the switch of the apparatus. This leads to a measurement dependent joint distribution $\mu_{x,y}$ for the distant apparatuses. Consequently, the corresponding classical probabilities are produced by
\begin{equation}\label{eq:prob1}
    P(a,b|x,y) = \int d\xi d\eta ~ \mu_{x,y}(\xi,\eta)f_x^a(\xi)g_y^b(\eta).
\end{equation}
We can further describe the effect of the operations as probability kernels~\cite{Spekken2005}, which transform any probability density $\mu(\xi)$ as
\begin{equation}
    \mu(\xi) \rightarrow \int d\xi' ~ T(\xi, \xi')\mu(\xi').
\end{equation}
Here the probability kernel $ T(\xi,\xi') \in [0,1]$ is normalized as $\int d\xi~T(\xi,\xi') = 1$ for any $\xi'$.

Since the two parties in the Bell experiment are causally disconnected, the measurement dependent distributions $\mu_{x,y}$ must be transformed from the original one by product probability kernels, that is,
\begin{equation}
    \mu_{x,y}(\xi,\eta) = \int d\xi' d\eta'~ T_x(\xi,\xi')T_y(\eta,\eta')\mu(\xi',\eta').
\end{equation}
This allows us to re-write the classical probabilities~(\ref{eq:prob1}) as
\begin{equation}\label{eq:prob2}
    P(a,b|x,y) = \int d\xi' d\eta' ~ \mu(\xi',\eta')\tilde{f}_x^a(\xi')\tilde{g}_y^b(\eta'),
\end{equation}
where the ``tilded'' function $\tilde{f}$ (similarly for $\tilde{g}$) is
\begin{equation}
    \tilde{f}_x^a(\xi') = \int d\xi ~ f_x^a(\xi) T_{x}(\xi,\xi'),
\end{equation}
which can be readily verified as a non-negative function bounded by unit, hence quantifies as a response function. However, we know that probabilities of form~(\ref{eq:prob2}) cannot recover those observed in a Bell test, hence neither do those in~(\ref{eq:prob1}). Therefore, classical descriptions for the statistics between the Bell measurement apparatuses are excluded.

\vspace{8pt}
Certainly, a classical determinist may contend that the nonlocal correlations must emanate from the microscopic Bell pair rather than from the objectively present apparatus in the macroscopic realm. This assertion align well with the acknowledgement of quantum physicists, who recognize that the distant apparatuses never become entangled; they are merely used to convey correlations within the underlying microscopic entangled particles.

However, the crux of the matter here does not rely on the origin of the observed nonlocal correlation. Rather, it resolves around the fundamental realization that the apparatuses in a Bell experiment defy classical descriptions. Essentially, when two systems become correlated, it becomes inconsistent to delineate one as quantum and the other as classical. Note that this conclusion is grounded in empirical evidence obtained from existing Bell experiments, in contrast to several no-go theorems~\cite{Sahoo2004mixing,Salcedo1996absence,Caro1999,Terno2006} for quantum-classical hybrid models, which depends on assumptions regarding specific forms of the underlying dynamics.

\vspace{8pt}
The currently scrutinized Bell test apparatuses are composed of quantum matters. To incorporate gravity, imagine a scenario where the Bell experiment interfaces with a Cavendish setup designed for gravitational force measurement. As illustrated in figure~\ref{fig:BellCavendish}, a Cavendish experiment involves two massive spheres affixed to a suspended rod. These spheres, in turn, exert gravitational attractions on two smaller balls that are independently suspended. The resultant torsion ensuring from the displacement of the smaller balls is measurable and serves to determine the gravitational forces.

In this integrated setup, each party in the Bell test can adjust the orientation of the rod attaching the larger spheres based on their respective Bell measurement outcomes, i.e., the values of $a$ and $b$. This rotational adjustment, either clockwise or counter-clockwise, creates distinct gravitational field configurations detectable by the responses of the inner smaller balls. Therefore, the integrated setup can be viewed again as a binary measurement scheme, with the smaller balls serving as the apparatus ``pointers''. The crucial difference, however, lies in the fact that the targeted object under scrutiny is the underlying gravitational field. Following the same argument elucidated earlier concerning macroscopic nonlocality, the observation of Bell correlations among the binary outcomes indicated by the smaller balls would rule out classical interpretations of the gravitational field. That is, a classical determinist attempting to replicate the observed correlations will fail at modeling the underlying gravitational fields using classical equations of motion, even with stochasticity in the classical phase space.

\vspace{8pt}
Integrating the classical Cavendish setup with a loophole-free Bell experiment~\cite{Hensen2015,Handsteiner2017,Storz2023} can be achieved relatively seamlessly. However, a noteworthy consideration is that the additional operations conducted on the massive objects will unfold over a much longer time scale than the actual quantum measurements performed in the original Bell experiment. To avoid locality loophole, the space separation between the distant parties in the Bell test must be significantly increased. For instance, an operational time window on the order of one second would necessitate entanglement distribution over the distance from the Earth to the Moon. This poses a significant experimental challenge. 

Modern technologies facilitating gravitational force measurement with significantly reduced masses~\cite{Rosi2014,Quinn2013,Westphal2021} may eventually lead to an experiment devoid of the locality loophole. Nevertheless, arguably, with mild assumptions, the strict requirement of spacelike separation in the Cavendish experiment may be relaxed. It suffices to maintain causal disconnection solely for the Bell experiment: once an outcomes signal is obtained from the Bell measurement and transmitted to the Cavendish setup, the sequential dynamics---such as the movements and responses of the massive spheres---are entirely determined by the local classical equations of motion, which are assumed not to be influenced by other distant operations. It is crucial to note that this condition does not contradict our overarching aim, that the gravitational field cannot be purely classical. The observed quantum effect is reflected in the nonlocal correlations between the distant fields. In a single observation of the Bell measurement, which fixes the initial condition of the Cavendish experiment, classical dynamics are not anticipated to break down. Therefore, established evidence for nonlocal correlations from the accomplished loophole-free Bell experiments~\cite{Hensen2015,Handsteiner2017,Storz2023}, together with the validity of classical dynamics for local macroscopic objects, significantly substantiated the quantum nature of gravity.

The examined Bell-Cavendish though experiment precludes the possibility of fundamentally characterizing gravitational field as a classical field. It is crucial to emphasize, however, that the affirmation of the quantum nature of gravity does not mandate the quantization of the gravitational field. Alternative perspectives, for instance, viewing gravity as an emerged phenomenon~\cite{Jacobson1995,Padmanabhan2010,Verlinde2011}, may still remain viable within the quantum framework. Implications of nonlocal correlations in gravity for these proposals deserve further exploration.

\vspace{8pt}
Acknowledgement.---This work was supported in part by the U.S. Department of Energy, Office of Science, Office of Advanced Scientific Computing Research, through the Quantum Internet to Accelerate Scientific Discovery Program, and in part by U.S. Department of Energy under the LDRD program at Los Alamos.

\bibliography{references}

\end{document}